\documentclass[10pt,letterpaper,twocolumn]{article} 

\usepackage{ol2}
\usepackage[draft]{hyperref}
\usepackage{amsmath}

\begin{document}
\bibliographystyle{ol}

\twocolumn[ 

\title{Generation of Bragg similaritons in nonlinear fiber Bragg gratings with gain}

\author{Martin Laprise, Michel Pich\'e, and R\'eal Vall\'ee}

\address{Centre d'Optique, Photonique et Lasers (COPL), Universit\'e Laval, Qu\'ebec, Canada, G1V 0A6}

\begin{abstract}We have explored through numerical simulations the amplification of laser pulses in nonlinear fiber Bragg gratings with gain and normal dispersion. The distortion of the temporal intensity profile caused by the higher-order dispersion coefficients is examined. The stabilizing effect of fourth-order dispersion on pulse propagation is demonstrated. For some conditions, the amplified pulse shape evolves towards a Bragg similariton, a quasi-parabolic temporal profile with wave-breaking free temporal characteristics.\end{abstract}


 ] 

\noindent 
The amplification of short laser pulses in the normal dispersion regime has been largely studied in recent years since it is a key consideration in order to achieve high-energy pulses from ytterbium-doped fiber laser systems.  It is now well known that high intensity pulse propagation in an optical fiber with normal group velocity dispersion (GVD) is subject to the optical wave-breaking instability.  The effects of this instability are known to be alleviated by the use of a parabolic pulse profile~\cite{anderson_parabolic}. This idea has been successfully extended to amplifiers with normal GVD by introducing the self-similarity concept~\cite{6701815,7200738}. It has been shown that the nonlinear Schr\"{o}dinger (NLS) equation with gain admits of a self-similar parabolic solution (or similariton) characterized by a parabolic temporal intensity profile with a linear chirp; the field distribution and the pulse spectrum evolve self-similarly during amplification, making pulse compression feasible even at high power.

Since similaritons are asymptotic solutions to the NLS equation with gain, any arbitrary input pulse will eventually evolve towards the parabolic pulse propagation regime. However, in practice, many constraints need to be fulfilled by the amplifier and the seed pulses in order to reach the parabolic regime in a finite length of amplifier. It has also been demonstrated that the effectiveness of the self-similar regime is inherently limited by the Raman response of the fiber material and the finite bandwidth of the amplifier~\cite{soh2006efp,soh2006efp2}. Therefore, in order to obtain the shortest pulses with the largest peak power, the length of the amplifier must be kept below a critical value. This constraint could be relaxed if one could design an amplifier with a larger value of normal GVD. One potential avenue to increase the effective GVD would consist in propagating the pulses through a fiber Bragg grating outside its photonic bandgap. In such a case, one could expect to mimic the classical parabolic pulse amplification regime with a much higher value of GVD.

In this letter, we present the results of a numerical investigation of pulse amplification in nonlinear fiber Bragg gratings (NLFBG) with gain. The pulses amplified in such media, termed Bragg similaritons (BS), have been found to share common properties with classical similaritons generated in standard optical fibers with gain. We will also show that the amplified pulses are sensitive to the large value of the third-order dispersion (TOD) induced by the fiber Bragg gratings. However, our numerical simulations reveal that the fourth-order dispersion (FOD) also present alleviates the effects of TOD which normally leads to a shock-wave type of instability. 

The propagation of an optical pulse in an NLFBG with gain can be described by the following pair of equations:
\begin{multline} \label{Eq1}
	i \frac{\partial u_f}{\partial z} + \frac{i}{v_g} \frac{\partial u_f}{\partial t} + \delta(z) u_f + \kappa(z) u_b \\
		+ \gamma \left(|u_f|^2 + 2|u_b|^2\right) u_f - \frac{i g}{2} u_f = 0
\end{multline}
\begin{multline} \label{Eq2}
	-i \frac{\partial u_b}{\partial z} + \frac{i}{v_g} \frac{\partial u_b}{\partial t} + \delta(z) u_b + \kappa(z) u_f \\
		+ \gamma \left(|u_b|^2 + 2|u_f|^2\right) u_b - \frac{i g}{2} u_b = 0
\end{multline}

where $u_f$ and $u_b$ represent the field envelopes of the electric field distributions corresponding to the forward and backward propagating waves, $v_g$ represents the group-velocity in the absence of the grating, $\gamma$ is the nonlinear coefficient, $\kappa(z)$ and $\delta(z)$ are the coupling and detuning coefficients of the grating. The constant $g$ represents the uniform small-signal gain of the amplifier. In an NLFBG without gain or losses, when the peak power of the pulse ($P_0$) is low enough to meet the condition $\gamma P_0 << \kappa$ and the TOD induced by the grating $\beta_3^g$ is negligible, we can reduce Eqs. (\ref{Eq1}-\ref{Eq2}) to a single effective NLS equation. If we consider the evolution of the pulse in an NLFBG amplifier in the absence of gain saturation and for pulses with spectral bandwidth less then the amplifier bandwidth, the pulse evolution can be described by an effective NLS equation with gain (NLSG)~\cite{PhysRevA.43.2467}:
\begin{equation} \label{Eq3}
	i\frac{\partial U}{\partial z} - \frac{\beta_2^g}{2} \frac{\partial^2 U}{\partial T} + \gamma_g |U|^2U - i\frac{g}{2} U= 0
\end{equation}

where $\beta_2^g$ and $\gamma_g$ are the effective GVD and the effective nonlinear parameter of the grating, $g$ the gain coefficient and $U$ is the Bloch-wave envelope. Since we know from previous analyses that the parabolic similariton is solution to this equation, we could expect the pulse evolution in NLFBGs with gain to share common properties with classical similaritons. The term Bragg similariton refers specifically to the self-similar solution of this effective NLSG equation.

\begin{figure}[htb]
\centerline{
\includegraphics[width=8.3cm]{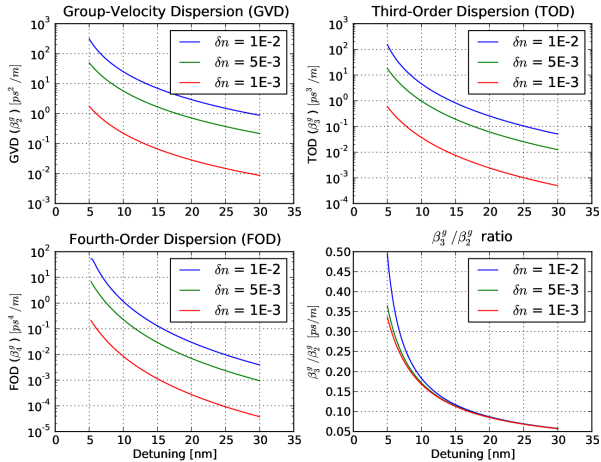}}
 \caption{Dispersion coefficients for an NLFBG in transmission near the bandgap for three different values of index modulation: $\delta n=10^{-2}$ (blue), $\delta n=5.0 \times 10^{-2}$ (green) and $\delta n=10^{-3}$ (red).}
	\label{fig1}
\end{figure}

In order to get a substantial increase of the effective GVD in an NLFBG operated in transmission, we need to propagate the pulse at a frequency very close to the photonic bandgap and/or use a grating with a very large index variation. In such cases, the effect of TOD cannot be neglected anymore and we have to evaluate how the BS is affected by higher-order dispersion. Fig.~\ref{fig1} shows the different dispersion coefficients for NLFBGs used in transmission. In these conditions, the GVD parameter ($\beta_2^g$) is many orders of magnitude higher than that of a classical fiber, but its TOD ($\beta_3^g$) is also very high. It is now well known that the presence of TOD, or more precisely a large $\beta_3/\beta_2$ ratio, has a detrimental effect on the self-similar regime~\cite{bale2010impact,zhang2009third}; hense we have to minimize this ratio in order to generate a BS. We can also take note that the effect of TOD will be minimized for a narrow pulse spectrum. The fourth-order dispersion ($\beta_4^g$) is also non-negligible and, as we will see, it has a significant effect on pulse propagation near the bandgap.

\begin{figure}[htb]
\centerline{
\includegraphics[width=7.0cm]{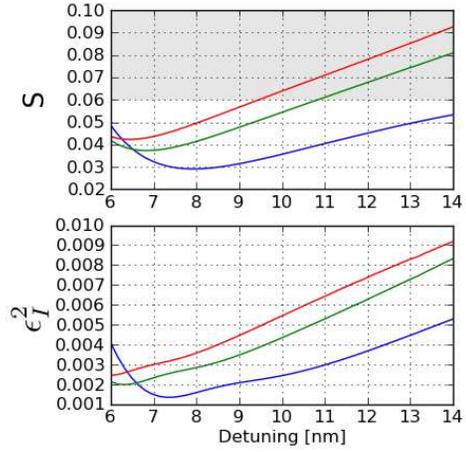}}
 \caption{Evolution of the skew parameter and parabolic misfit with detuning for an input gaussian pulse of 1.0 (blue), 3.5 (green) and 6.0 nJ (red).}
	\label{fig2}
\end{figure}

Following the notation of Bale et al.~\cite{bale2010impact} we characterized the effect of the higher-order dispersion coefficients on the propagated pulses in the NLFGB with the skew parameter which quantifies the asymmetry of the pulse temporal shape:

\begin{equation} \label{Eq3}
	S = \int{A^2 \xi^3 d\xi}
\end{equation}

where $\xi$ is a normalized time variable and A the normalized amplitude. The skew parameter has been computed for different values of detuning by propagating a 5.0-ps gaussian pulse over a distance of 1~m with a classical Split-Step Fourier (SSF) algorithm with the corresponding higher-order dispersion ($\beta_{n<5}$) and the following NLFBG parameters: $\delta n=1.0 \times 10^{-2}$, g~=~29 dB/m, $\gamma$ = 54 $km^{-1}$ $W^{-1}$. The misfit parameter between the pulse intensity profile $|U|^2$ and the parabolic fit $|U_{fit}|^2$ has also been computed for the same conditions:

\begin{equation} \label{Eq4}
	\epsilon_I^{2} = \int{\left( |U|^2 - |U_{fit}|^2 \right)^2} dt /  \int{|U|^4} dt
\end{equation}

\begin{figure}[htb]
\centerline{
\includegraphics[width=8.3cm]{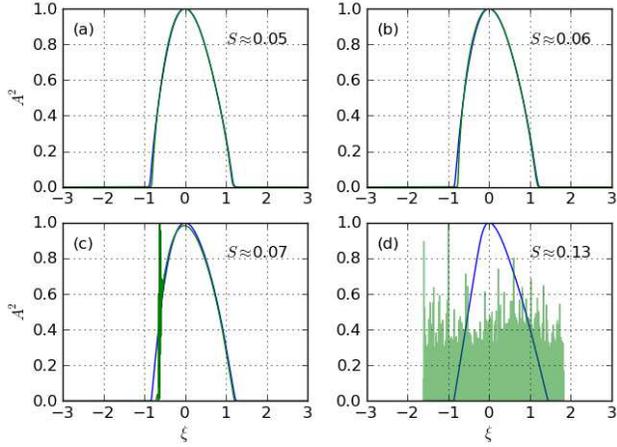}}
 \caption{Effect of fourth-order dispersion on the wave breaking on an asymmetric pulse for different values of skew parameter: $S\approx0.05$~(a), $S\approx0.06$~(b), $S\approx0.07$~(c) and $S\approx0.13$~(d), with (blue) and without FOD (green).}
	\label{fig3}
\end{figure}

Fig. \ref{fig2} shows the skew and the misfit parameters ($\epsilon_I^{2}$) for various values of detuning and initial pulse energy. Interestingly, if we are assuming that these two parameters are good indicators of the departure of the pulse from the classical similariton, this departure increases with detuning even if the $\beta_3^g/\beta_2^g$ ratio is decreasing. In the present case, the optimal detuning for the formation of a similariton-like pulse is between 6-8 nm. The gray region in the upper graph of Fig.~\ref{fig2} shows the condition for which a shock-wave type of instability normally starts to develop ($S > 0.06$) in the absence of FOD. However, in the present case, the large value of  FOD inhibits the formation of instabilities, even in the case of a strongly distorted pulse. Fig.~\ref{fig3} shows some examples of classical similaritons distorted by TOD with different skew parameters with and without FOD. The pulse has been propagated over four  values of distance with a classical SSF algorithm using the dispersion parameters $\beta_2^g=24.7~ps^2/m$ and $\beta_3^g=4.5~ps^3/m$. We can clearly see the onset of wave breaking on the left side of the pulse propagating without FOD when, for the same conditions, the pulse propagating with $\beta_4^g=1.1~ps^4/m$ is immune to this kind of instability. 

\begin{figure}[htb]
\centerline{
\includegraphics[width=8.3cm]{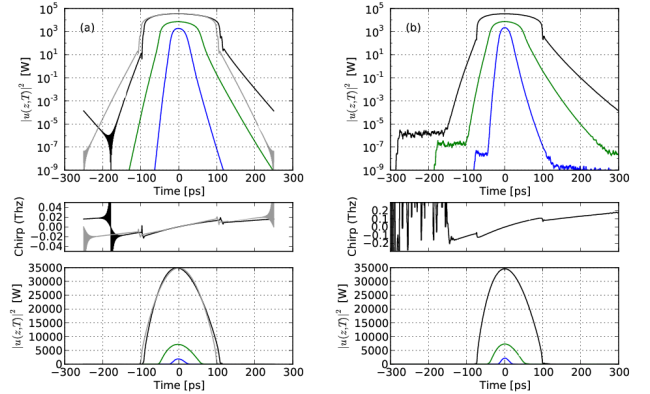}}
 \caption{Temporal profile after 1635~(blue), 3272~(green) and 4906~ps~(black) of propagation in the NLFBG using a classical SSF algorithm~(a) and the OSSM~(b). The gray curve in (a) corresponds to a classical similariton without TOD nor FOD.}
	\label{fig4}
\end{figure}

Considering the optimal conditions previously found, we simulated pulse propagation in an NLFBG with gain by solving numerically Eqs.~(\ref{Eq1}-\ref{Eq2}) to all orders of dispersion with the Optimized Split-Step Method (OSSM) described in~\cite{toroker2008oss}.  The NLFBG used in these simulations has a length of 1~m, a constant gain coefficient $g$=29 dB/m, a coupling constant $\kappa$=24.5 $cm^{-1}$ and a detuning constant $\delta$=-60.5 $cm^{-1}$. The dispersion coefficients are given by: $\beta_2^g=83.4~ps^2/m$, $\beta_3^g=23.9~ps^3/m$ and $\beta_4^g=9.8~ps^4/m$. The effective refractive index in the absence of the grating was equal to n=1.45 and the nonlinear coefficient equal to $\gamma=54~km^{-1} W^{-1}$. The input gaussian pulse has a temporal width of 5.0~ps and a peak power of 1.13~kW. In order to compare the results with propagation in a simple medium, the simulations have also been repeated with the classical SSF algorithm with the corresponding dispersion and nonlinear coefficients. Fig.~\ref{fig4} shows the temporal pulse characteristics on linear and logarithmic scales from both methods. Since the OSSM is a temporal method and the group velocity in the NLFBG is changing with detuning, the three temporal profiles are identified by their propagation time. The profiles obtained with the OSSM and the classical SSF algorithm are very similar. We can suppose that the minor differences between the two curves originate from the cumulative effects of the higher-order dispersion coefficients ($\beta_{n>4}$) which were left out in the SSF simulations. More interestingly, the pulses share many properties with classical similaritons. The central part of the pulse evolves from a gaussian profile to a slightly skewed parabolic shape. When plotted on a logarithmic scale, the intensity profile presents abruptly decreasing wings typical of compact parabolic similaritons. The intensity profile also presents exponentially decaying low-intensity wings typical of the intermediate asymptotic regime of similaritons. Finally, the chirp in the high intensity region of the pulse exhibits an excellent linearity.

In summary, we have numerically investigated the optimal condition for the amplification of a pulse in an NLFBG with gain in the normal dispersion regime. The stabilizing effet of FOD on the similariton propagation in the presence of TOD has been demonstrated. Finally, we have shown that pulses amplified in an NLFBG share common properties with classical similaritons generated in standard optical fibers with gain. This solution could open the possibility of parabolic amplification at wavelengths usually confined into the anomalous dispersion regime such as Er-doped fiber amplifiers.

This work was supported by the Canadian Institute for Photonic Innovations and National Sciences and Engineering Research Council of Canada. M. Laprise e-mail address is martin.laprise.1@ulaval.ca.

\pagebreak


\begin{thebibliography}{99}
\bibitem{anderson_parabolic}
D.~Anderson, M.~Desaix, M.~Karlsson, M.~Lisak, and M.~L. Quiroga-Teixeiro, "Wave-breaking-free pulses in nonlinear-optical fibers", J. Opt. Soc. Am. B \textbf{10}, 1185 (1993).

\bibitem{6701815}
M.~E. Fermann, V.~I. Kruglov, B.~C. Thomsen, J.~M. Dudley, and J.~D. Harvey, "Self-Similar Propagation and Amplification of Parabolic Pulses in Optical Fibers", Phys. Rev. Lett. \textbf{84}, 6010 (2000).

\bibitem{7200738}
V.~I. Kruglov, A.~C. Peacock, J.~D. Harvey, and J.~M. Dudley, "Self-similar propagation of parabolic pulses in normal-dispersion fiber amplifiers", J. Opt. Soc. Am. B \textbf{19}, 461 (2002).

\bibitem{soh2006efp}
D.~B. Soh, J.~Nilsson, and A.~B. Grudinin, "Efficient femtosecond pulse generation using a parabolic amplifier combined with a pulse compressor. II. Finite gain-bandwidth effect", J. Opt. Soc. Am. B \textbf{23}, 10 (2006).

\bibitem{soh2006efp2}
D.~B. Soh, J.~Nilsson, and A.~B. Grudinin, "Efficient femtosecond pulse generation using a parabolic amplifier combined with a pulse compressor. I. Stimulated Raman-scattering effects", J. Opt. Soc. Am. B \textbf{23}, 1
  (2006).

\bibitem{bernier2009ytterbium}
M.~Bernier, R.~Vall\'{e}e, B.~Morasse, C.~Desrosiers, A.~Saliminia, and
  Y.~Sheng, "Ytterbium fiber laser based on first-order fiber Bragg gratings written with 400nm femtosecond pulses and a phase-mask", Opt. Express \textbf{17}, 18887 (2009).

\bibitem{PhysRevA.43.2467}
C.~Martijn~de Sterke and J.~E. Sipe, "Gap solitary waves with gain and loss", Phys. Rev. A \textbf{43}, 2467 (1991).

\bibitem{bale2010impact}
B.~Bale and S.~Boscolo, "Impact of third-order fibre dispersion on the evolution of parabolic optical pulses", Journal of Optics \textbf{12}, 015202 (2010).

\bibitem{zhang2009third}
S.~Zhang, G.~Zhao, A.~Luo, and Z.~Zhang, "Third-order dispersion role on parabolic pulse propagation in dispersion-decreasing fiber with normal group-velocity dispersion", Applied Physics B: Lasers and Optics \textbf{94}, 227 (2009).

\bibitem{toroker2008oss}
Z.~Toroker and M.~Horowitz, "Optimized split-step method for modeling nonlinear pulse propagation in fiber Bragg gratings", J. Opt. Soc. Am. B \textbf{25}, 448 (2008).


\end{thebibliography}
\end{document}